\documentclass[a4paper,11pt]{article}
\usepackage[dvips]{graphics}
\usepackage[utf8]{inputenc}
\usepackage[T1]{fontenc}
\usepackage{cmap}
\usepackage{jcappub}

\def\m{\overline{m}}
\def\a{\widetilde{\alpha}}

\title{(In)stability of black holes in the $4D$ Einstein-Gauss-Bonnet and Einstein-Lovelock gravities}

\author[\star,\dagger]{R. A. Konoplya}%
\author{and}%
\author[\star,\diamond]{A. Zhidenko}%

\affiliation[\star]{Institute of Physics and Research Centre of Theoretical Physics and Astrophysics,\\ Faculty of Philosophy and Science, Silesian University in Opava,\\ Bezručovo nám. 13, CZ-746 01 Opava, Czech Republic}

\affiliation[\dagger]{Peoples Friendship University of Russia (RUDN University),\\ 6 Miklukho-Maklaya Street, Moscow 117198, Russian Federation}

\affiliation[\diamond]{Centro de Matemática, Computação e Cognição (CMCC),\\ Universidade Federal do ABC (UFABC),\\ Rua Abolição, CEP: 09210-180, Santo André, SP, Brazil}

\emailAdd{roman.konoplya@gmail.com}
\emailAdd{olexandr.zhydenko@ufabc.edu.br}

\arxivnumber{2003.12492}

\abstract{
A $(3+1)$-dimensional Einstein-Gauss-Bonnet effective description of gravity has been recently formulated as the $D \to 4$ limit of the higher dimensional field equations after the rescaling of the coupling constant. This approach has been recently extended to the four-dimensional Einstein-Lovelock gravity. Although validity of the regularization procedure has not been shown for the general case, but only for a wide class of metrics, the black-hole solution obtained as a result of such a regularization is also an exact solution in the well defined $4D$ Einstein-Gauss-Bonnet theory suggested by Aoki, Gorji and Mukohyama [arXiv:2005.03859] and in the scalar-tensor effective classical theories.  Here we study the eikonal gravitational instability of asymptotically flat, de Sitter and anti-de Sitter black holes in the four dimensional Einstein-Gauss-Bonnet and Einstein-Lovelock theories. We find parametric regions of the eikonal instability for various orders of the Lovelock gravity, values of coupling and cosmological constants, and share the code which allows one to construct the instability region for an arbitrary set of parameters. For the four-dimensional Gauss-Bonnet black holes we obtain the region of stability in analytic form. Unlike the higher dimensional Einstein-Lovelock case, the eikonal instability serves as an effective cut-off of higher curvature Lovelock terms  for the $4D$ black holes.}

\begin{document}
\maketitle

\section{Introduction}
Stability of a black-hole metric against small (linear) perturbations of spacetime is the necessary condition for viability of the black-hole model under consideration. Therefore, a number of black-hole solutions in various alternative theories of gravity were tested for stability \cite{Konoplya:2011qq}. One of the most promising approaches to construction of alternative theories of gravity is related to the modification of the gravitational sector via adding higher curvature corrections to the Einstein action. This is well motivated by the low energy limit of string theory. Among higher curvature corrections, the Gauss-Bonnet term (quadratic in curvature) and its natural generalization to higher orders of curvature in the Lovelock form \cite{Lovelock:1971yv,Lovelock:1972vz} play an important role.

The Lovelock theorem states that only metric tensor and the Einstein tensor are divergence free, symmetric, and concomitant of the metric tensor and its derivatives in four dimensions \cite{Lovelock:1971yv,Lovelock:1972vz}. Therefore, it was concluded that the appropriate vacuum equations in $D=4$ are the Einstein equations (with the cosmological term). In $D>4$ the theory of gravity is generalized by adding higher curvature Lovelock terms to the Einstein action.

Black hole in the $D>4$ Einstein-Gauss-Bonnet gravity and its Lovelock generalization were extensively studied and various peculiar properties were observed. For example, the life-time of the black hole whose geometry is only slightly corrected by the Gauss-Bonnet term  is  characterized by a much longer lifetime and a few orders smaller evaporation rate \cite{Konoplya:2010vz}. The eikonal quasinormal modes in the gravitational channel break down the correspondence between the eikonal quasinormal modes and null geodesics \cite{Cardoso:2008bp,Konoplya:2017wot}. However, apparently the most interesting feature of higher curvature corrected black hole is the gravitational instability: When the coupling constants are not small enough, the black holes are unstable and the instability develops at high multipoles numbers   \cite{Dotti:2005sq,Gleiser:2005ra,Konoplya:2008ix,Takahashi:2012np,Takahashi:2011qda,Yoshida:2015vua,Cuyubamba:2016cug,Konoplya:2017ymp,Konoplya:2017lhs,Konoplya:2017zwo}. Therefore, it was called the \emph{eikonal instability} \cite{Cuyubamba:2016cug}.

Recently, it was claimed that there was found the way to bypass the Lovelock's theorem \cite{Glavan:2019inb} by performing a kind of dimensional regularization of the Gauss-Bonnet equations and obtaining of a four-dimensional metric theory of gravity with diffeomorphism invariance and second order equations of motion.
The approach was first formulated in $D > 4$ dimensions and then, the four-dimensional theory is defined as the limit $D \to 4$ of the higher-dimensional theory after the rescaling of the coupling constant $\alpha \to \alpha/(D-4)$. The properties of black holes in this theory, such as (in)stability, quasinormal modes and shadows, were considered in \cite{Konoplya:2020bxa}, while the innermost circular orbits were analyzed in \cite{Guo:2020zmf}. The generalization to the charged black holes and an asymptotically anti-de Sitter and de Sitter cases in the $4D$ Einstein-Gauss-Bonnet theory was considered in \cite{Fernandes:2020rpa} and to the higher curvature corrections, that is, the $4D$ Einstein-Lovelock theory, in \cite{Konoplya:2020qqh,Casalino:2020kbt}. Some further properties of black holes  for this novel theory, such as axial symmetry and thermodynamics, were considered in \cite{Wei:2020ght,Kumar:2020owy,Hegde:2020xlv,Zhang:2020qew,Ghosh:2020syx,Zhang:2020qam,Kumar:2020uyz,HosseiniMansoori:2020yfj,Wei:2020poh,Singh:2020nwo,Churilova:2020aca,Mishra:2020gce,Heydari-Fard:2020sib,Konoplya:2020cbv,Jin:2020emq,Zhang:2020sjh,EslamPanah:2020hoj,NaveenaKumara:2020rmi,Aragon:2020qdc,Malafarina:2020pvl,Yang:2020czk,Cuyubamba:2020moe,Mahapatra:2020rds,Shu:2020cjw,Casalino:2020pyv,Liu:2020evp,Devi:2020uac,Ma:2020ufk,Liu:2020yhu,Kumar:2020sag,Churilova:2020mif,Ge:2020tid,Zeng:2020dco,Ghosh:2020cob,Yang:2020jno}.

It should be pointed out that such a ``naïve'' formulation of the dimensional regularization has faced some criticism, starting from a straightforward observation of the lack of the tensorial description for the corresponding theory \cite{Gurses:2020ofy}. It has been found further that in some cases different ways for regularization lead to nonuniqueness of some solutions, such as Taub-NUT black holes \cite{Hennigar:2020lsl}. It was pointed out that in four dimensions there is no four-point graviton scattering tree amplitudes other than those leading to the Einstein theory, so that additional degrees of freedom, for instance, a scalar field $(\partial\phi)^4$, should be added for consistency \cite{Bonifacio:2020vbk}. In addition, the nonlinear perturbations of the metric cannot be regularized by taking the limit $D \to 4$ due to divergent terms appearing in the corresponding equations of the Gauss-Bonnet theory \cite{Arrechea:2020evj}.

In order to solve the above problems additional scalar degrees of freedom were proposed in \cite{Lu:2020iav,Kobayashi:2020wqy} through a Kaluza-Klein reduction of a $D$-dimensional theory, which in the limit $D\to4$ leads to a particular subclass of the Horndeski theory with a scalar field $(\partial\phi)^4$. An alternative approach for introducing the scalar field, which does not exploit a particular assumption on the extra-dimensional geometry, leading to the same scalar-tensor theory (when the internal Kaluza-Klein space is flat), has been proposed in \cite{Fernandes:2020nbq,Hennigar:2020fkv}. The theory admits two vacua, one corresponding to the Einstein gravity and the other one -- to the regularized Gauss-Bonnet case, and do not have an additional propagating degree of freedom associated with the scalar field \cite{Lu:2020mjp}. Thus, in order to study gravitational dynamics, we need to take into account only gravitational degrees of freedom. However, it turns out that the gravitational degrees of freedom in such a scalar-tensor theory are infinitely
strong coupled due to lack of the quadratic kinetic term of the scalar field \cite{Kobayashi:2020wqy}.

A consistent description for the theory has been given in \cite{Aoki:2020lig}, where, using the ADM decomposition, it was shown that the regularization either
\begin{itemize}
\item breaks the diffeomorphism invariance, leading to a particular vacuum and implying no scalar-field degree of freedom,\\
or,
\item introduces an extra degree of freedom given by a scalar field, which is in agreement with the Lovelock theorem.
\end{itemize}

Here we study the linear stability of asymptotically flat, de Sitter and anti-de Sitter $4D$ Einstein-Lovelock black holes.
We show that not only maximally symmetric spacetimes, but even much less symmetric time-dependent linear perturbations can be studied in four dimensions with arbitrary Lovelock couplings by taking the limit $D \to 4$ without encountering divergences.
We find the parametric regions of the eikonal instability for the $4D$ Einstein-Gauss-Bonnet-(anti-)de Sitter black holes and various examples for its Lovelock extension. {We show that the positive values of coupling constants are bounded by the instability region. Thus, small black holes are not allowed in the $4D$ Einstein-Lovelock description with positive couplings: When the coupling constants are sufficiently large compared to the black-hole size, such black holes are always eikonally unstable. This situation is qualitatively different from the higher dimensional Einstein-Lovelock theory, where the effect of higher curvature terms cannot be discarded, because there the instability still allows for large values of the coupling constants. Inequalities determining the instability region for the four-dimensional Gauss-Bonnet black holes are derived.

Our paper is organized as follows. In Sec.~\ref{sec:blackhole} we briefly describe the static $4D$ Einstein-Lovelock black hole solution. Sec.~\ref{sec:perturbations} is devoted to the gravitational perturbations of the black holes, and Sec.~\ref{sec:instability} discusses their eikonal instability. Finally, in Conclusions, we summarize the obtained results.

\section{Static black holes in the four-dimensional Lovelock theory}\label{sec:blackhole}
The Lagrangian density of the Einstein-Lovelock theory has the form
\cite{Lovelock:1971yv}:
\begin{eqnarray}\label{Lagrangian}
  \mathcal{L} &=& -2\Lambda+\sum_{m=1}^{\m}\frac{1}{2^m}\frac{\alpha_m}{m}
  \delta^{\mu_1\nu_1\mu_2\nu_2 \ldots\mu_m\nu_m}_{\lambda_1\sigma_1\lambda_2\sigma_2\ldots\lambda_m\sigma_m}\,
  R_{\mu_1\nu_1}^{\phantom{\mu_1\nu_1}\lambda_1\sigma_1} R_{\mu_2\nu_2}^{\phantom{\mu_2\nu_2}\lambda_2\sigma_2} \ldots R_{\mu_m\nu_m}^{\phantom{\mu_m\nu_m}\lambda_m\sigma_m},
\end{eqnarray}
where $\delta^{\mu_1\mu_2\ldots\mu_p}_{\nu_1\nu_2\ldots\nu_p}$
is the generalized totally antisymmetric Kronecker delta, $R_{\mu\nu}^{\phantom{{\mu\nu}}\lambda\sigma}$ is the Riemann tensor, $\alpha_1=1/8\pi G=1$ and $\alpha_2,\alpha_3,\alpha_4,\ldots$ are arbitrary constants of the theory.

The Euler-Lagrange equations, corresponding to the Lagrangian density (\ref{Lagrangian}) read \cite{Kofinas:2007ns}:
\begin{eqnarray}
 \Lambda\delta^{\mu}_{\nu} &=& R^{\mu}_{\nu}-\frac{R}{2}\delta^{\mu}_{\nu}+\sum_{m=2}^{\m}\frac{1}{2^{m+1}}\frac{\alpha_m}{m}
 \delta^{\mu\mu_1\nu_1 \ldots\mu_m\nu_m}_{\nu\lambda_1\sigma_1\ldots\lambda_m\sigma_m}
R_{\mu_1\nu_1}^{\phantom{\mu_1\nu_1}\lambda_1\sigma_1} 
\ldots R_{\mu_m\nu_m}^{\phantom{\mu_m\nu_m}\lambda_m\sigma_m}\,.
\label{Lovelock}
\end{eqnarray}

The antisymmetric tensor is nonzero only when the indices $\mu,\mu_1,\nu_1,\ldots\mu_m,\nu_m$ are all distinct. Thus, the general Lovelock theory is such that $2\m <D$. In particular, for $D=4$, we have $\m=1$ corresponding to the Einstein theory \cite{Lovelock:1972vz}. When $D=5$ or $6$, $\m=2$ and one has the (quadratic in curvature) Einstein-Gauss-Bonnet theory with the coupling constant $\alpha_2$.

Following \cite{Konoplya:2017lhs}, we introduce
\begin{equation}\label{amdef}
\a_m=\frac{\alpha_m}{m}\frac{(D-3)!}{(D-2m-1)!}=\frac{\alpha_m}{m}\prod_{p=1}^{2m-2}(D-2-p)
\end{equation}
and consider the limit $D\to 4$ while $\a_m$ remain constant. In this way, we obtain the regularized $4D$ Einstein-Lovelock theory formulated in \cite{Konoplya:2020qqh}, which generalizes the approach of \cite{Glavan:2019inb} used for the Einstein-Gauss-Bonnet theory. In the Einstein-Gauss-Bonnet case ($\m=2$) the above equation reads
\begin{equation}
\alpha_2 = \frac{2 \a_2}{(D-3)(D-4)}.
\end{equation}
Then, taking the limit $D\to 4$ we see that
\begin{equation}
\alpha_2 \to \frac{2 \a_2}{D-4}.
\end{equation}
Notice that our units differ by a factor of $2$ from those used in \cite{Glavan:2019inb} and coincide with the units of \cite{Fernandes:2020rpa}. Prior to \cite{Glavan:2019inb} the dimensional regularization of the Einstein-Gauss-Bonnet theory was suggested by Y.~Tomozawa~\cite{Tomozawa:2011gp}.

Although the Lagrangian (\ref{Lagrangian}) diverges in the limit $D \to 4$, no singular terms appear in the Einstein-Lovelock equations for any $D\geq3$.
In particular, following \cite{Konoplya:2020qqh} one can find the four-dimensional static and spherically symmetric metric, described by the metric
\begin{equation}\label{Lmetric}
  ds^2=-f(r)dt^2+\frac{1}{f(r)}dr^2 + r^2 (d\theta^2+\sin^2\theta d\phi^2).
\end{equation}

The metric function $f(r)$ is defined through a new variable $\psi(r)$,
\begin{equation}\label{Lfdef}
f(r)=1-r^2\,\psi(r),
\end{equation}
which satisfies the algebraic equation
\begin{equation}\label{MEq}
W[\psi(r)]\equiv\psi(r)+\sum_{m=2}^{\m}\a_m\psi(r)^m-\frac{\Lambda}{3}=\frac{2M}{r^{3}}\,,
\end{equation}
where $M$ is the asymptotic mass \cite{Myers:1988ze}.

The Gauss-Bonnet theory ($\m=2$) leads to the two branches \cite{Fernandes:2020rpa}:
\begin{equation}
  f(r)=1-\frac{r^2}{2\a_2}\left(-1\pm\sqrt{1+4\a_2\left(\frac{2M}{r^3}+\frac{\Lambda}{3}\right)}\right),
\end{equation}
one of which, corresponding to the ``+'' sign, is perturbative in $\a_2$, while for the ``-'' the metric function $f(r)$ goes to infinity when $\a_2 \to 0$. Notice that the above solution was also obtained in \cite{Cai:2009ua,Cognola} in a different context, when discussing quantum correction to entropy.

The higher-order Lovelock corrections result in more branches, only one of which is perturbative in $\a_m$. Following \cite{Konoplya:2017lhs}, we consider here only the perturbative branch, so that we recover the Einstein theory \cite{Tangherlini:1963bw} in the limit $\a_m\to0$. In particular, for $\m=3$ and $\a_3\geq\a_2^2/3$,
\begin{equation}
f(r)=1-\frac{\a_2r^2}{3\a_3}\left(A_+(r)-A_-(r)-1\right),
\end{equation}
where
$$
A_{\pm}(r)=\sqrt[3]{\sqrt{F(r)^2+\left(\frac{3\a_3}{\a_2^2}-1\right)^3}\pm F(r)},\qquad
F(r)=\frac{27\a_3^2}{2\a_2^3}\left(\frac{2M}{r^3}+\frac{\Lambda}{3}\right)+\frac{9\a_3}{2\a_2^2}-1\,.
$$

It is convenient to measure all dimensional quantities in units of the horizon radius $r_H$. For the asymptotic mass we obtain
\begin{equation}\label{Mdef}
  2M=r_H\left(1+\sum_{m=2}^{\m}\frac{\a_m}{r_H^{2m-2}}-\frac{\Lambda r_H^2}{3}\right).
\end{equation}

For $\Lambda>0$ the perturbative branch is asymptotically de Sitter, so that $\Lambda$ can be expressed in terms of the de Sitter horizon $r_C$ as follows
\begin{eqnarray}\label{LdSdef}
  \frac{\Lambda}{3}&=&\frac{1}{r_C^2+r_Cr_H+r_H^2}+\sum_{m=2}^{\m}\a_m \frac{r_C^{3-2m}-r_H^{3-2m}}{r_C^3-r_H^3}.
\end{eqnarray}

When $\Lambda<0$, we introduce the AdS radius $R$, by assuming that the metric function has the following asymptotic $f(r)\to r^2/R^2$ as $r\to\infty$, and the cosmological constant is given by
\begin{equation}\label{LAdS}
\frac{\Lambda}{3}=-\frac{1}{R^2}+\sum_{m=2}^{\m}\frac{(-1)^m\a_m}{R^{2m}}.
\end{equation}

The metric function $f(r)$ for the perturbative branch of the general Einstein-Lovelock black hole can be obtained numerically \cite{Konoplya:2020qqh}.\footnote{The Mathematica\textregistered{} code for the metric-function calculation is available from \url{https://arxiv.org/src/2003.07788/anc/LovelockBH.nb}.}

\section{Gravitational perturbations}\label{sec:perturbations}
Following \cite{Takahashi:2010}, we consider linear perturbations of the $D$-dimensional spherically symmetric black hole, which we separate into tensor, vector, and scalar channels according to their transformations respectively the rotation group on a $(D-2)$-sphere:
\begin{itemize}
\item  Although for $D>4$ the \emph{tensor-type perturbations} have dynamic degrees of freedom, in the limit $D\to4$ these perturbations are pure gauge.
\item  For the \emph{vector-type (axial) perturbations} we choose the Regge-Wheeler gauge, so that the nonzero perturbations of the metric tensor are
\begin{subequations}
\begin{eqnarray}
  \delta g_{ti}&=&\delta g_{it}=v(t,r)V_i,\\
  \delta g_{ri}&=&\delta g_{ir}=w(t,r)V_i,
\end{eqnarray}
\end{subequations}
where $i,j=2,3,\ldots(D-1)$ are indices of the $D-2$ sphere and the vector harmonics $V_i$ depend on the corresponding coordinate and satisfy
    \begin{equation}\label{vectorharmonics}
      \nabla_iV^i=0, \qquad \nabla^j\nabla_jV^i=V^i-\ell(\ell+D-3)V^i.
    \end{equation}
\item For the \emph{scalar-type (polar) perturbations} we use the Zerilli gauge
\begin{subequations}
\begin{eqnarray}
\delta g_{tt}&=&g_{tt}H_0(t,r)S,\\
\delta g_{tr}=\delta g_{rt}&=&H_1(t,r)S,\\
\delta g_{rr}&=&g_{rr}H_2(t,r)S,\\
\delta g_{ij}&=&g_{ij}K(t,r)S,
\end{eqnarray}
\end{subequations}
 where the scalar harmonics $S$ obey
    \begin{equation}\label{scalarharmonics}
      \nabla^j\nabla_jS=-\ell(\ell+D-3)S.
    \end{equation}
\end{itemize}
The integer number $\ell=2,3,4,\ldots$ is called the multipole number.

Substituting perturbed metric
$$g_{\mu\nu}\rightarrow g_{\mu\nu}+\delta g_{\mu\nu}$$
into (\ref{Lovelock}) and neglecting higher orders of $\delta g_{\mu\nu}$, one can obtain the following equations \cite{Takahashi:2010}
\begin{subequations}
\begin{eqnarray}
\nonumber
-\frac{1}{2r^D}\left(T'(r)v(t,r)\left(\frac{\nabla^j\nabla_jV^i}{D-3}+V^i\right)+f(r)\frac{\partial}{\partial r}T(r)\left(r\frac{\partial v}{\partial r}-r\frac{\partial w}{\partial t}-2v(t,r)\right)V^i\right)=0\\
(\mu=i, \nu=0),\qquad\\
\nonumber
-\frac{1}{2r^D}\left(T'(r)w(t,r)\left(\frac{\nabla^j\nabla_jV^i}{D-3}+V^i\right)+\frac{T(r)}{f(r)}\frac{\partial}{\partial t}\left(r\frac{\partial v}{\partial r}-r\frac{\partial w}{\partial t}-2v(t,r)\right)V^i\right)=0\\
(\mu=i, \nu=r),\qquad\\
\nonumber
-\frac{1}{2(D-3)r^D}\left(\frac{T'(r)}{f(r)}\frac{\partial v}{\partial t}-\frac{\partial}{\partial r}\left(f(r)T'(r)w(t,r)\right)\right)\left(\nabla_i V_j+\nabla_j V_i\right)=0\\
(\mu=i, \nu=j),\qquad
\end{eqnarray}
\end{subequations}
for the vector-type perturbations.
For the scalar-type perturbations we have \cite{Takahashi:2010}
\begin{subequations}
\begin{eqnarray}
\nonumber
(T(r)H_2(t,r)+rT'(r)K(t,r))\nabla^j\nabla_j S-(D-2)r\frac{\partial}{\partial r}f(r)T(r)H_2(t,r)S\qquad\qquad\\
\nonumber
+(D-2)rT'(r)K(t,r)S+\frac{D-2}{2}r^2T(r)\frac{\partial K}{\partial r}S+(D-2)f(r)\frac{\partial}{\partial r}r^2T(r)\frac{\partial K}{\partial r}S=0\\
(\mu=0, \nu=0),\qquad\\
\nonumber
H_1(t,r)\nabla^j\nabla_j S+(D-2)r\left(K(t,r)-H_2(t,r)+r\frac{\partial K}{\partial r}-\frac{rf'(r)}{2f(r)}K(t,r)\right)S=0,\\
(\mu=1, \nu=0),\qquad\\
\nonumber
(T(r)H_0(t,r)+rT'(r)K(t,r))\nabla^j\nabla_j S+(D-2)r\Biggl(2T(r)\frac{\partial H_1}{\partial t}-\frac{rT(r)}{f(r)}\frac{\partial^2 K}{\partial t^2}+T'(r)K(t,r)\\
\nonumber
+r\left(f(r)T'(r)+\frac{f'(r)T(r)}{2}\right)\frac{\partial K}{\partial r}-(f(r)T'(r)+f'(r)T(r))H_2(t,r)+f(r)T(r)H_0(t,r)\Biggr)S\\
(\mu=1, \nu=1),\qquad
\end{eqnarray}
\begin{eqnarray}
\nonumber
\Biggl(T'(r)H(t,r)-rT'(r)\frac{\partial K}{\partial r}+\frac{f'(r)T(r)}{2f(r)}\left(H_2(t,r)-H_0(t,r)\right)\qquad\qquad\qquad\qquad\qquad\qquad\\
-T(r)\frac{\partial H_0}{\partial r}+\frac{T(r)}{r}H_0(t,r)+\frac{T(r)}{f(r)}\frac{\partial H_1}{\partial t}\Biggr)\nabla_iS=0\qquad
(\mu=i, \nu=1),\qquad\\
\left(rT''(r)K(t,r)+T'(r)H_0(t,r)+T'(r)H_2(t,r)\right)\nabla_i\nabla_jS=0\qquad
(\mu=i, \nu=j),\qquad
\end{eqnarray}
\end{subequations}
where we introduced the function
\begin{equation}
T(r)\equiv r^{D-3} W'[\psi(r)]=r^{D-3}\left(1+\sum_{m=2}^{\m}m\a_m\psi(r)^{m-1}\right).
\end{equation}
It essential that all the above equations are well defined in the limit $D\to4$.

In \cite{Takahashi:2010} it was shown that, after introducing the functions $\Psi_v(t,r)$ and $\Psi_s(t,r)$ as
\begin{eqnarray}
w(t,r)&=&\frac{r\Psi_v(t,r)}{f(r)\sqrt{|T'(r)|}},
\\
K(t,r)&=&\frac{rf(r)\left(\dfrac{\partial\Psi_s}{\partial r}+\dfrac{T'(r)}{T(r)}\Psi_s(t,r)\right)-\dfrac{\ell(\ell+D-3)}{D-2}\Psi_s(t,r)}{f(r)-\dfrac{r}{2}f'(r)+\dfrac{\ell(\ell+D-3)}{D-2}},
\end{eqnarray}
it is possible to reduce the perturbation equations to the wave-like form:
\begin{equation}\label{wavelike}
\left(\frac{\partial^2}{\partial t^2}-\frac{\partial^2}{\partial r_*^2}+V_i(r_*)\right)\Psi_{i}(t,r_*)=0,
\end{equation}
where $r_*$ is the tortoise coordinate,
\begin{equation}
dr_*\equiv \frac{dr}{f(r)}=\frac{dr}{1-r^2\psi(r)},
\end{equation}
and $i$ stands $v$ (\emph{vector}) and $s$ (\emph{scalar}) types of gravitational perturbations.
The wave-like equations (\ref{wavelike}) describe the perturbation dynamics since $v(t,r)$ and $w(t,r)$ depend on $\Psi_v(t,r)$ and the functions $H_0(t,r)$, $H_1(t,r)$, $H_2(t,r)$, and $K(t,r)$ depend on $\Psi_s(t,r)$ and its derivatives.

After taking the limit $D\to4$ the effective potentials $V_s(r)$ and $V_v(r)$ are given by the following expressions:
\begin{eqnarray}
V_v(r)&=&\frac{(\ell-1)(\ell+2)}{rT(r)}\frac{d}{dr_*}T(r)+R(r)\frac{d^2}{dr_*^2}\Biggr(\frac{1}{R(r)}\Biggr),\\
V_s(r)&=&\frac{\ell(\ell+1)}{rP(r)}\frac{d}{dr_*}P(r)+\frac{P(r)}{r}\frac{d^2}{dr_*^2}\left(\frac{r}{P(r)}\right),
\end{eqnarray}
where
$$
R(r)=r\sqrt{|T'(r)|},\qquad
P(r)=\frac{2(\ell-1)(\ell+2)-2r^3\psi'(r)}{\sqrt{|T'(r)|}}T(r).
$$

Notice that, since we consider the perturbative branch of solutions, then we have $T(r)>0$ for $r>r_H$ \cite{Konoplya:2017lhs}.

Although we can formally obtain expressions for the effective potentials when $T'(r)\leq0$, the kinetic term of perturbations in such points has a wrong (negative) sign. In this case the perturbations are linearly unstable, and this phenomenon was called the \emph{ghost} instability \cite{Takahashi:2010}.

\section{Eikonal (in)stability}\label{sec:instability}
Usually, we used to believe that if a gravitational instability takes place, it happens at the lowest $\ell=2$ multipole, while higher multipoles increase the centrifugal part of the effective potential and make the potential barrier higher, so that, usually,, higher $\ell$ are more stable. The eikonal instability we observe here is qualitatively different: higher $\ell$ leads not only to the higher height of the barrier, but also increases the depth of the negative gap near the event horizon. Then, at some sufficiently large $\ell$ the negative gap becomes so deep, that the bound state with negative energy becomes possible, which signifies the onset of instability.

For large $\ell$ the effective potential for the vector-type perturbations
\begin{equation}
V_{v} = \ell^2 \left(\frac{f(r)T'(r)}{rT(r)} + {\cal O}\left(\frac{1}{\ell}\right)\right)
\end{equation}
becomes positive-definite in the parametric regime, which is free from the ghost instability.

Therefore, the eikonal instability exists only in the scalar channel. For large $\ell$ the effective potential reads
\begin{equation}
V_{s}=\ell^2 \left(\frac{f(r)(2T'(r)^2-T(r)T''(r))}{2rT'(r)T(r)} + {\cal O}\left(\frac{1}{\ell}\right)\right),
\end{equation}
giving the following instability condition \cite{Takahashi:2010},
\begin{equation}\label{instability-cond}
2T'(r)^2-T(r)T''(r)<0.
\end{equation}

This condition can be test for each black-hole configuration. In the Einstein-Gauss-Bonnet case ($\m=2$) it sufficient to test if (\ref{instability-cond}) is satisfied at $r=r_H$ \cite{Konoplya:2017ymp}. Therefore, the problem is reduced to the polynomial inequality of fourth order in $\a_2$, which can be solved analytically. We find that the eikonal instability occurs if
\begin{equation}\label{GBinstability}
\frac{\a_2}{r_H^2}>\frac{\sqrt{6\sqrt{3}-10+\lambda^2}+\lambda}{2},\qquad \lambda=(2\sqrt{3}-3)\Lambda r_H^2-1.
\end{equation}

In the asymptotically flat case ($\Lambda=0$, $\lambda=-1$), the black hole has the eikonal instability in the scalar sector for
\begin{equation}
\frac{\a_2}{r_H^2}>\frac{\sqrt{6\sqrt{3}-9}-1}{2}\approx0.09.
\end{equation}
It is possible to show that the ghost instability always takes place for larger values of $\a_2$.

\begin{figure*}
\centerline{\resizebox{\linewidth}{!}{\includegraphics*{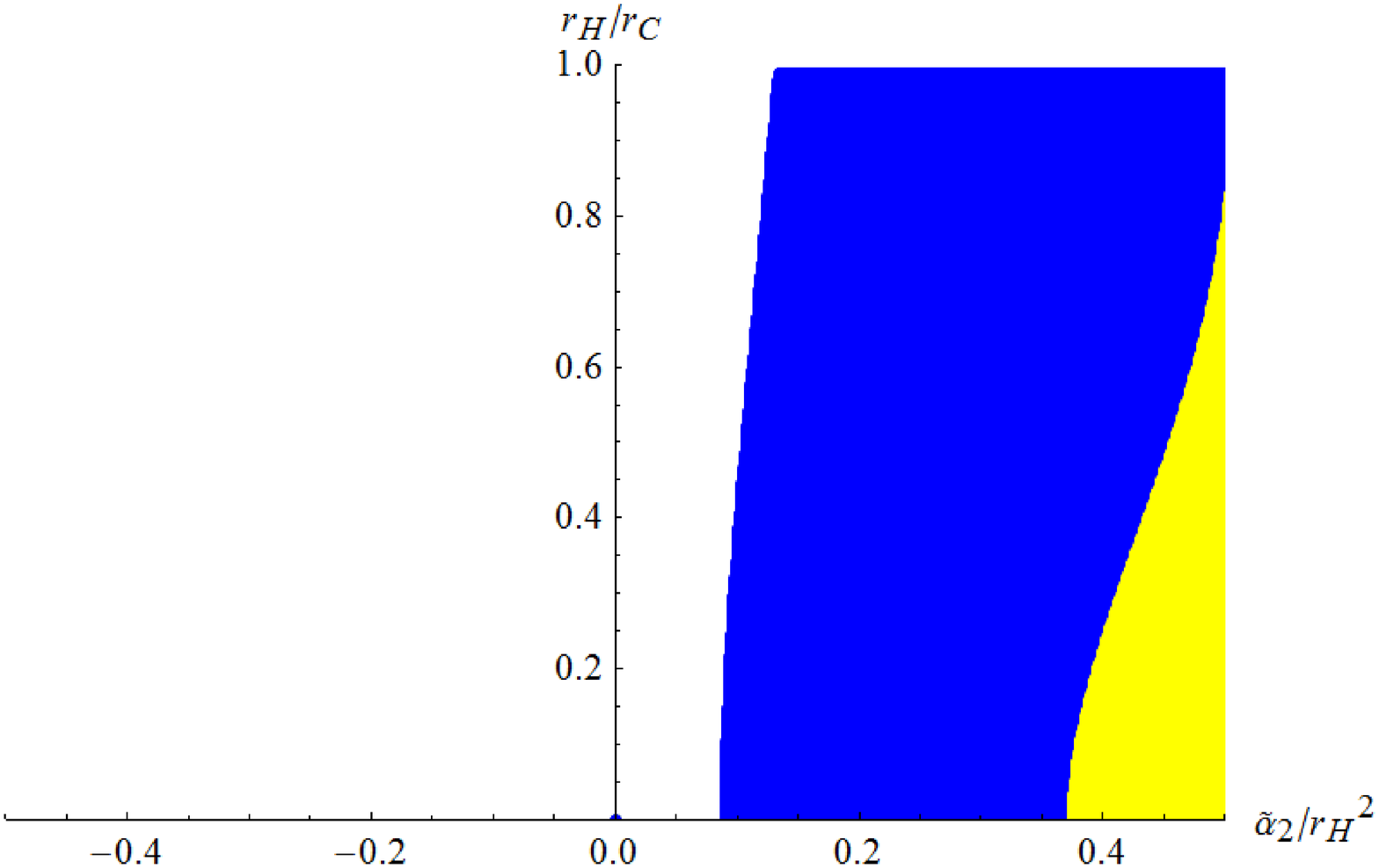}\includegraphics*{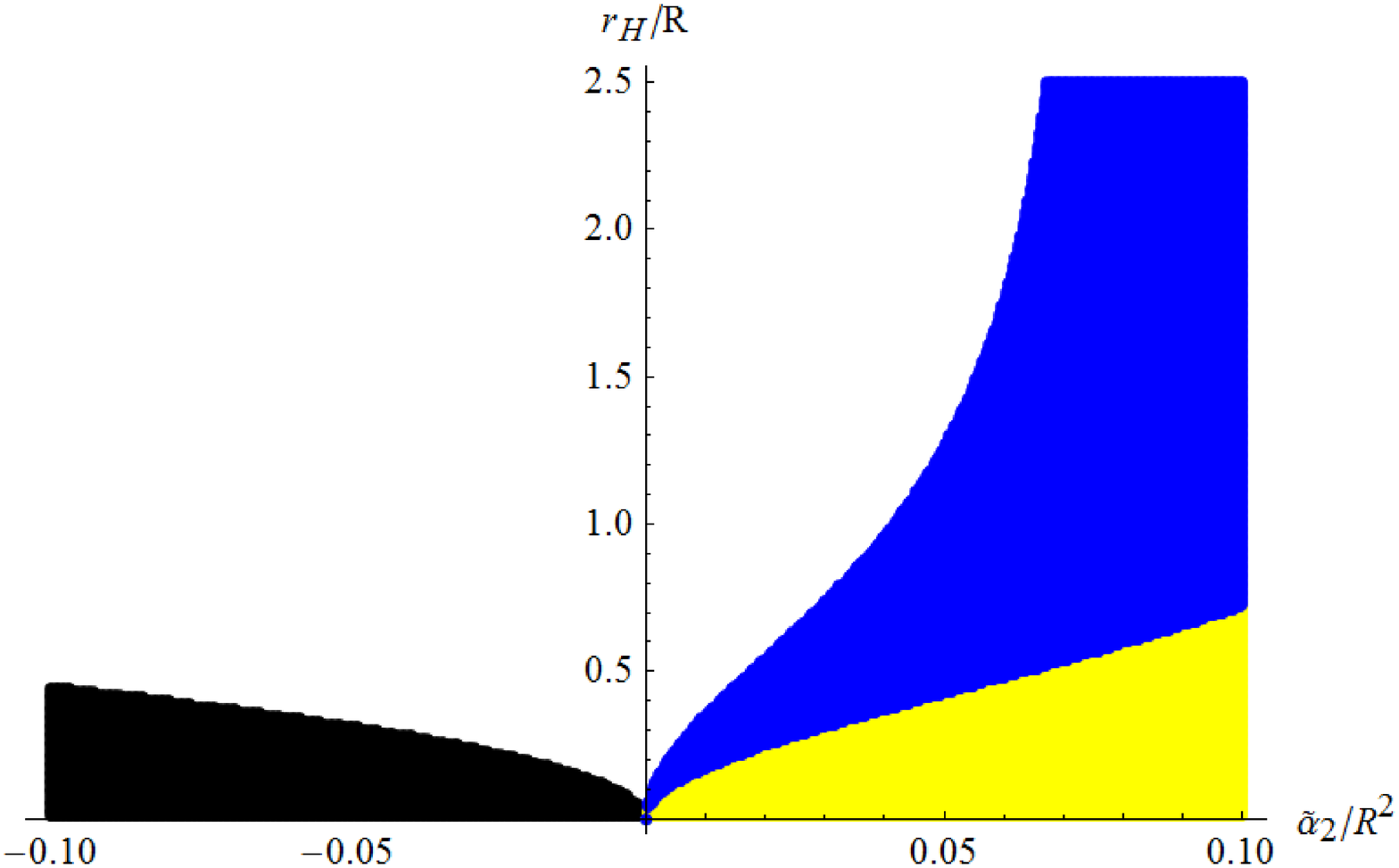}}}
\caption{The parametric region of the eikonal instability for the asymptotically de Sitter (left panel) and anti-de Sitter (right panel) four-dimensional Einstein-Gauss-Bonnet black holes. The black color is the excluded parametric region ($T(r_H)<0$), the yellow color is for the ghost instability ($T'(r_H)<0$), and the blue color is for the eikonal instability in the scalar sector.}\label{fig:Gauss-Bonnet}
\end{figure*}

By substituting (\ref{LdSdef}) and (\ref{LAdS}) into (\ref{GBinstability}) one can find the instability condition in the geometrized units. We show the parametric region of the eikonal instability on Fig.~\ref{fig:Gauss-Bonnet}. We see that the asymptotically flat and (anti-)de Sitter black holes in the Einstein-Gauss-Bonnet theory are stable for the whole range of valid parameters when $\alpha$ is negative and for sufficiently small values of positive $\alpha$.

For the higher-order Einstein-Lovelock theory it is not sufficient to test (\ref{instability-cond}) in the point $r=r_H$. In order to obtain the region of instability we have used the Wolfram Mathematica\textregistered{} code developed in \cite{Konoplya:2017lhs}.\footnote{The Mathematica\textregistered{} code for testing stability of the $4D$ Einstein-Lovelock black hole is available from \url{https://arxiv.org/src/2003.12492v1/anc/Lovelock-Stability.nb}.}

\begin{figure*}
\centerline{\resizebox{\linewidth}{!}{\includegraphics*{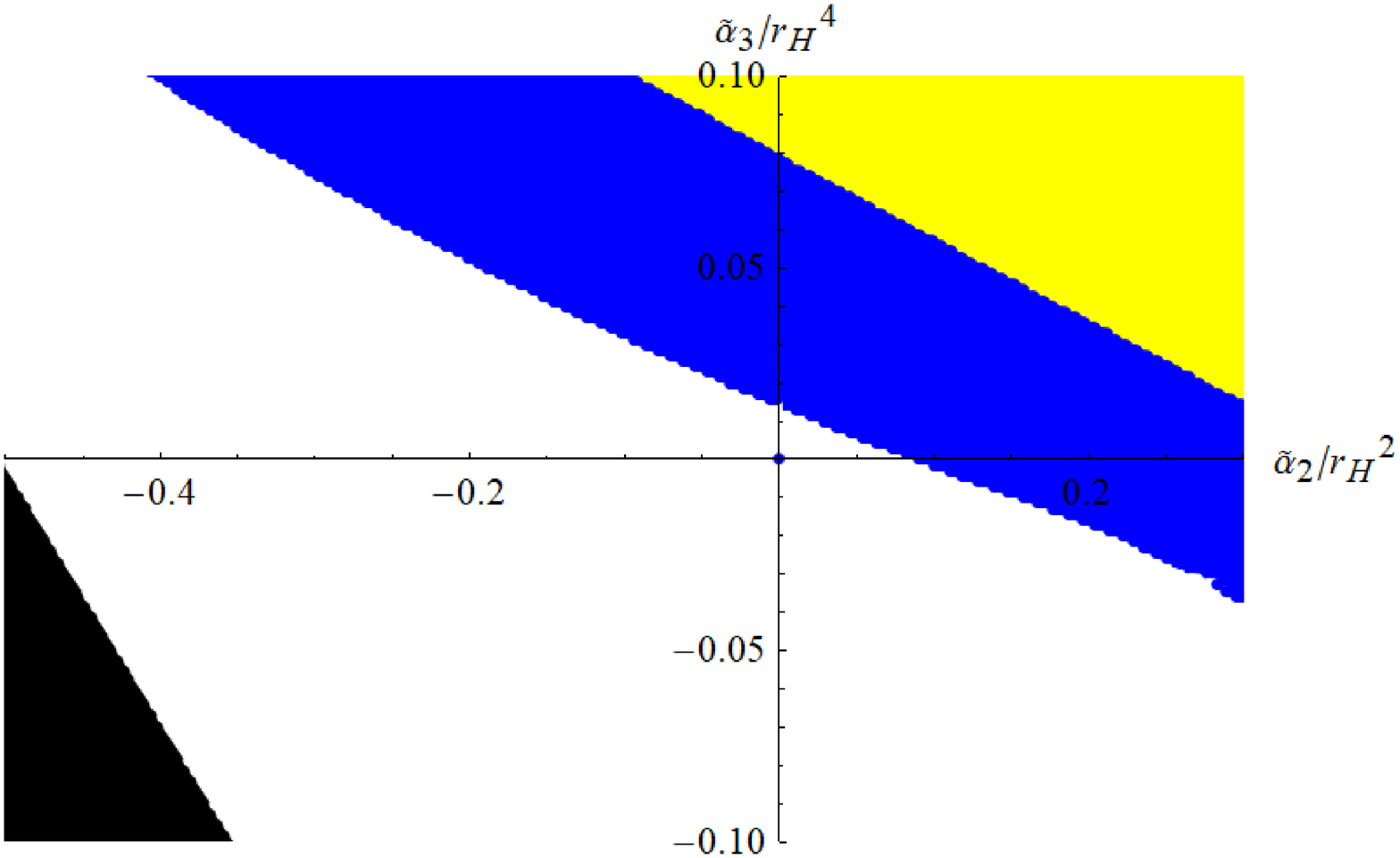}\includegraphics*{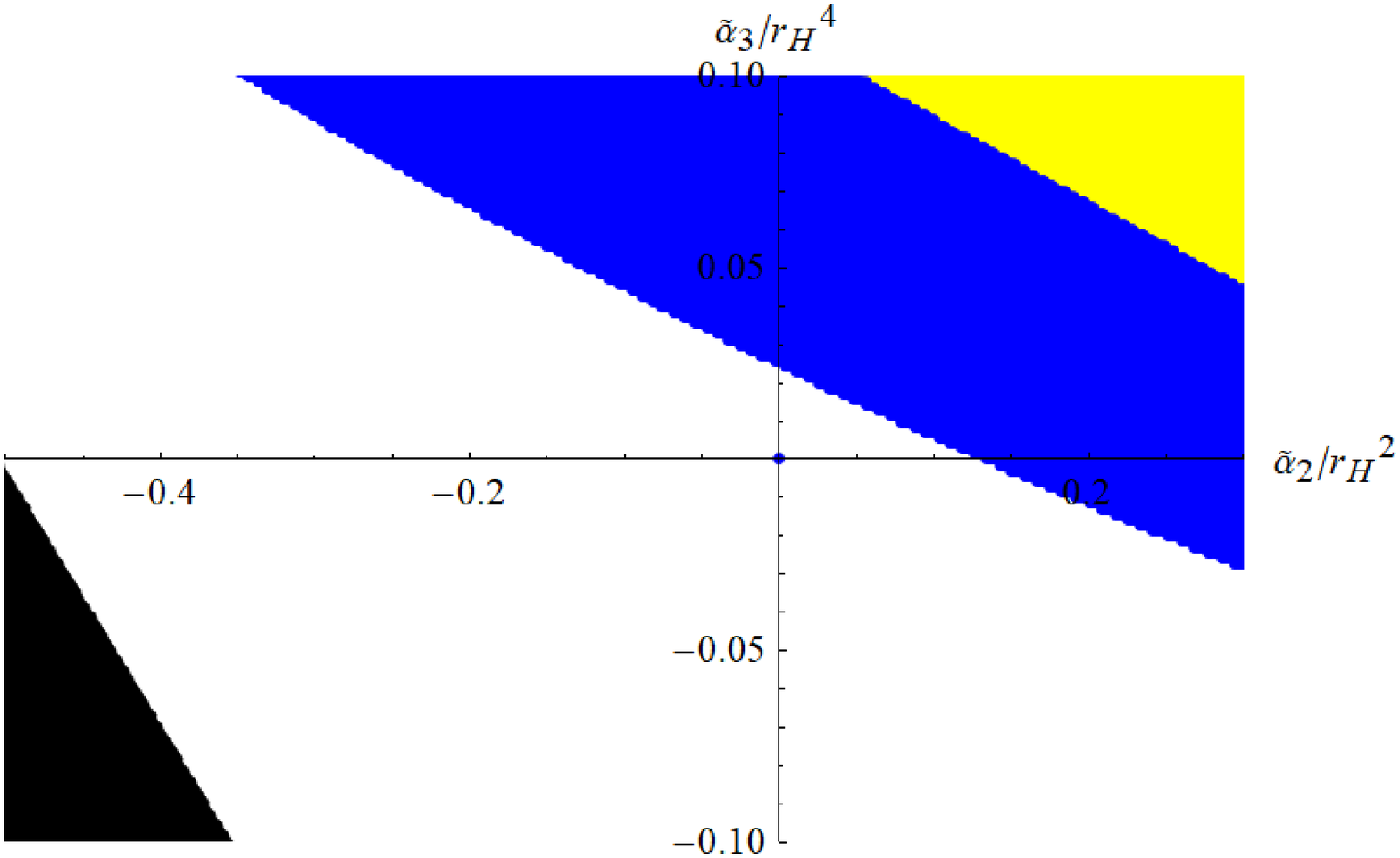}}}
\caption{The parametric region of the eikonal instability on the ($\a_2$,$\a_3$)-plane for the asymptotically flat, $\Lambda=0$, (left panel) and near extreme de Sitter, $r_{H}/r_{C} = 0.99$, (right panel) four-dimensional black holes in the Einstein-Lovelock theory of the third order in curvature. The black color is the excluded parametric region, the yellow color is for the ghost instability, and the blue color is for the eikonal instability in the scalar sector.}\label{Lovelock-dS}
\end{figure*}

\begin{figure*}
\centerline{\resizebox{\linewidth}{!}{\includegraphics*{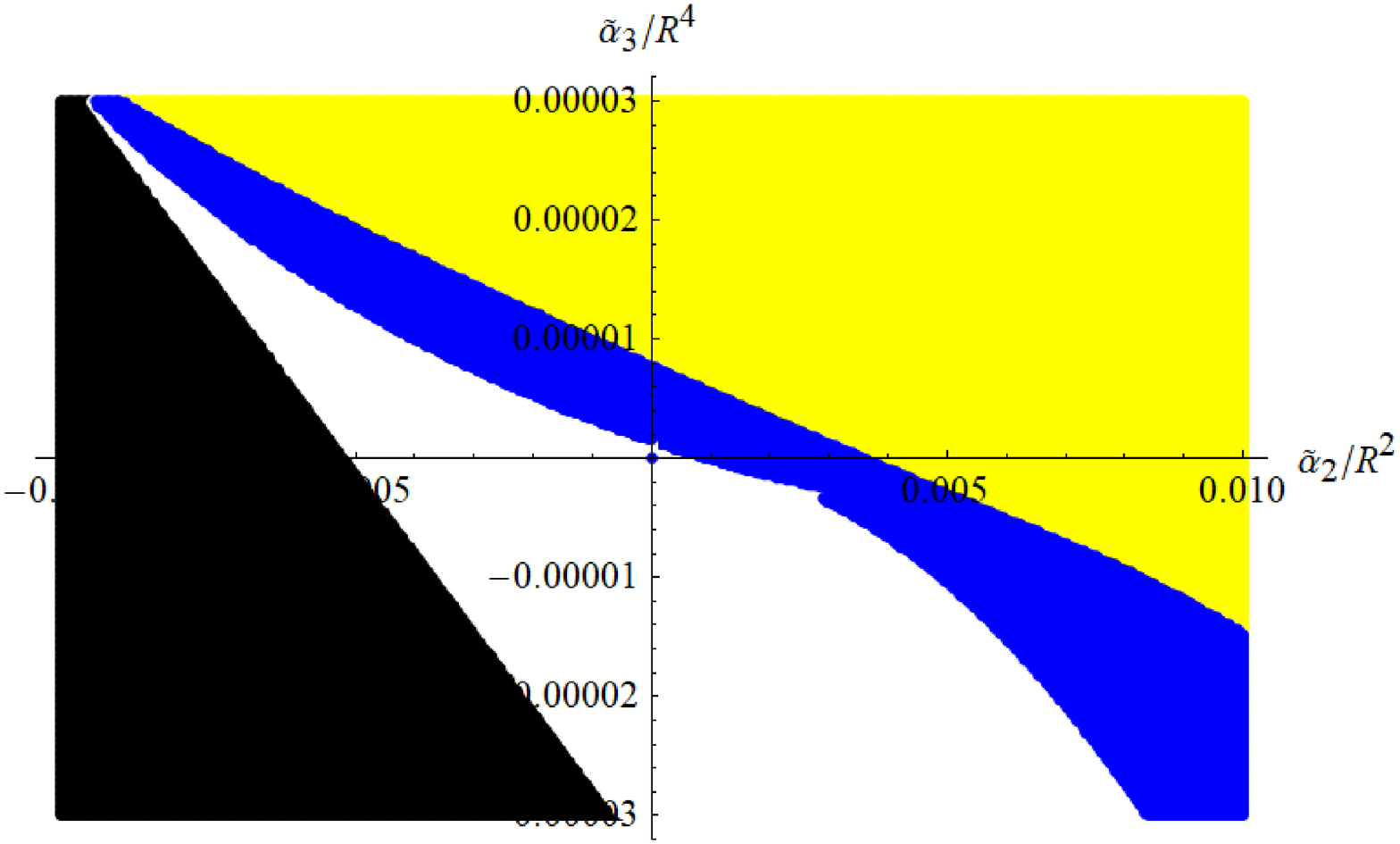}\includegraphics*{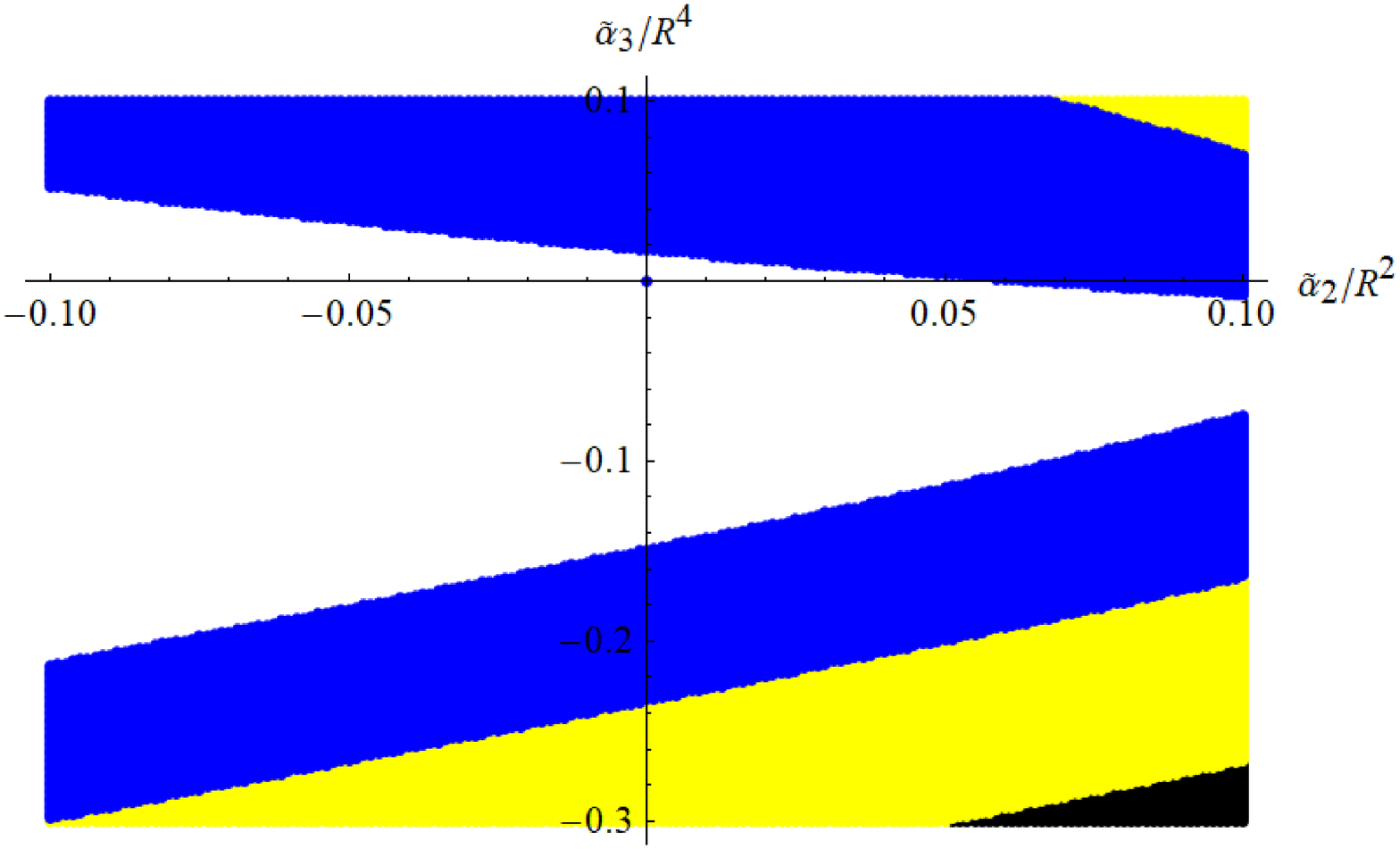}\includegraphics*{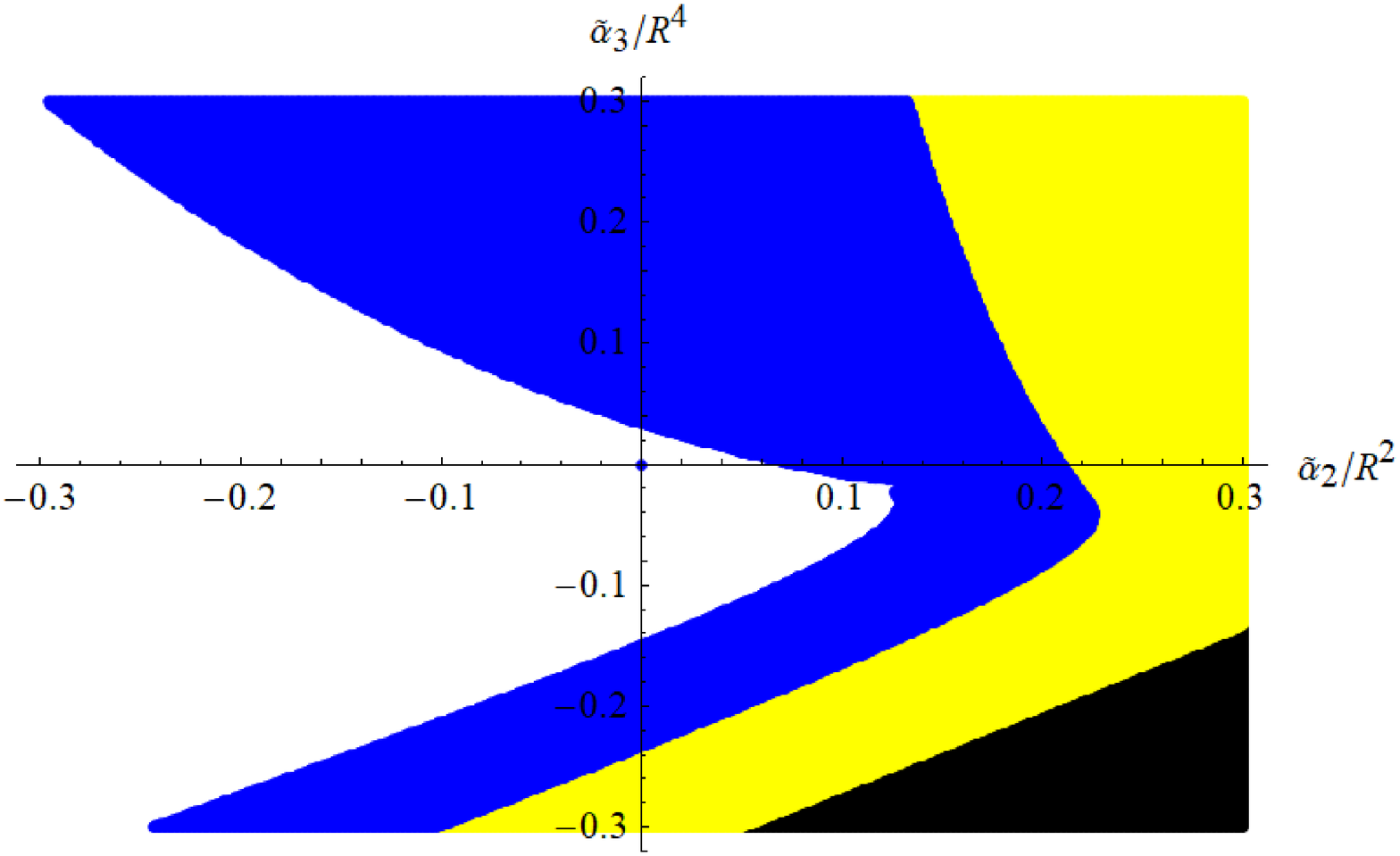}}}
\caption{The parametric region of the eikonal instability on the ($\a_2$,$\a_3$)-plane for the asymptotically anti-de Sitter black hole in the Einstein-Lovelock theory of the third order in curvature. From left to right: small black holes ($r_H/R=0.1$), medium black holes ($r_H/R=2$), and large black holes ($r_H/R=10$). The black color is the excluded parametric region, the yellow color is for the ghost instability, and the blue color is for the eikonal instability in the scalar sector.}\label{Lovelock-AdS}
\end{figure*}

From Figs.~\ref{Lovelock-dS} we see that when the Lovelock series is truncated at the third order ($\m=3$) the positive cosmological constant modifies the (in)stability region relatively softly. As for the four-dimensional Einstein-Gauss-Bonnet black holes, we see the $\Lambda$-term slightly increases the region of stability. For the asymptotically AdS black holes (Figs.~\ref{Lovelock-AdS}) we see that the region of stability shrinks as we decrease their size in units of the AdS radius $R$.

It is interesting to note that for the four-dimensional black holes the ghost instability occurs for $\a_3<\a_2^2/3$ and scalar-type eikonal instability exists for $\a_3<0$. Unlike in higher dimensional Lovelock theory, spherically symmetric black holes in $4D$ are always unstable for sufficiently large positive values of the coupling constants. In this way, the eikonal instability is an effective cut-off for the physically relevant black-hole solutions with positive couplings.

\section{Conclusions}
Here we analyzed the (in)stability of the asymptotically flat, de Sitter and anti-de Sitter $4D$ Einstein-Lovelock black holes. First of all, we showed that not only the background spherically symmetric solution can be regularized, but the higher dimensional time-dependent perturbation equations allow for the same dimensional regularization, showing no divergences in the limit $D\to4$. We showed that for all types of asymptotics the black holes are unstable unless the coupling constants are sufficiently small. Negative coupling constants allow for a much larger parametric region of stability. It is interesting that when limited by positive coupling constants, the eikonal instability strongly constrains values of the coupling constants at higher orders and serves as an effective cut-off of the series. This phenomena does not take place in the higher dimensional case, and, therefore it would be tempting to learn whether this cut-off due to the instability takes place in the well-defined and non-contradictory Einstein-dilaton-Lovelock theory.
For the general case of the Einstein-Lovelock theory of arbitrary order and a number of particular examples the regions of instability are found as algebraic inequalities. In a similar way, our work could be extended to the case of a charged black hole.

It is important to stress out that we have studied linear stability of spherically symmetric black holes within the \emph{classical theory of gravity}, obtained by regularizing the corresponding solutions of $D$-dimensional Einstein-Gauss-Bonnet \cite{Glavan:2019inb} or Einstein-Lovelock \cite{Konoplya:2020qqh,Casalino:2020kbt} theories. Recently it was shown that the simplest solutions of this class, dimensionally regularized Einstein-Gauss-Bonnet black holes, are also solutions of a subclass of the scalar-tensor Horndeski theories. These solutions are unambiguous since the Weyl part of the corresponding equations vanishes,}
\begin{equation}
C^{\mu\rho\lambda\sigma}C_{\nu\rho\lambda\sigma}-\frac{1}{4}\delta^{\mu}_{\nu}C^{\tau\rho\lambda\sigma}C_{\tau\rho\lambda\sigma}=0,
\end{equation}
where $C_{\nu\rho\lambda\sigma}$ is the Weyl tensor. As the scalar field does not add propagating degrees of freedom and the Weyl terms do not appear in the linear perturbation equations, our stability analysis has to be compatible with these theories.

Yet, the second-order perturbation equations cannot be consistently derived by taking the limit $D\to4$ \cite{Arrechea:2020evj} and the analysis of four-point graviton scattering amplitudes shows that the strongly-coupled scalar field must be taken into account at this order \cite{Bonifacio:2020vbk}. In this sense the regularized Gauss-Bonnet theory considered here is only a low-energy effective description of the scalar-tensor theory. The consistent Hamiltonian theory implies that the gravitational perturbations gain a correction to the dispersion relation in the ultraviolet regime due to counter terms, appearing in order to cancel divergences of the Weyl pieces \cite{Aoki:2020lig}. Although the linear perturbation equations for the full theory have not been derived yet, we notice that the Weyl tensor on the $(D-2)$-sphere does not appear in scalar-type and vector-type perturbation equations \cite{Kodama:2003kk}, indicating that the counter terms do not change the angular parts of the equations, which were used for our eikonal stability analysis. It is worth mentioning here that although the eikonal instability of the Gauss-Bonnet black holes manifests itself first at large multipole number, the similar unstable behavior is observed for lower $\ell$ as well \cite{Konoplya:2008ix}, corresponding to the low-energy perturbations at the threshold of instability, for which the ultraviolet corrections can be neglected.

Thus we conclude that the linear and higher-order perturbation analysis in the full theory could further limit the parametric region of stability of the $4D$-Gauss-Bonnet black holes. However, it is unlikely that the linearly unstable black holes, discussed here, can be stabilized when taking into account the strong scalar-field coupling.

As to the $4D$-Lovelock black holes, the consistent scalar-tensor theory has yet to be formulated. Nevertheless, the Kaluza-Klein reduction allows one to obtain higher-order Lovelock terms of the corresponding scalar-tensor theory \cite{Kobayashi:2020wqy}. It was shown in \cite{Kobayashi:2020wqy} that the scalar-tensor theory with the Lovelock term of order $\m$ yields the scalar-field term $(\partial\phi)^{2\m}$, confirming that the contributions from the higher-order Lovelock terms are high-energy correction to gravity.

\begin{acknowledgments}
The authors acknowledge the support of the grant 19-03950S of Czech Science Foundation (GAČR). This publication has been prepared with partial support of the ``RUDN University Program 5-100'' (R. K.).
\end{acknowledgments}


\begin{thebibliography}{99}
\bibitem{Konoplya:2011qq}
  R.~A.~Konoplya and A.~Zhidenko,
  Rev.\ Mod.\ Phys.\  {\bf 83}, 793 (2011)
  doi:10.1103/RevModPhys.83.793
  [arXiv:1102.4014 [gr-qc]].

\bibitem{Lovelock:1971yv}
  D.~Lovelock,
  J.\ Math.\ Phys.\  {\bf 12}, 498 (1971)
  doi:10.1063/1.1665613.

\bibitem{Lovelock:1972vz}
  D.~Lovelock,
  J.\ Math.\ Phys.\  {\bf 13}, 874 (1972)
  doi:10.1063/1.1666069.

\bibitem{Konoplya:2010vz}
  R.~A.~Konoplya and A.~Zhidenko,
  Phys.\ Rev.\ D {\bf 82}, 084003 (2010)
  doi:10.1103/PhysRevD.82.084003
  [arXiv:1004.3772 [hep-th]].

\bibitem{Cardoso:2008bp}
  V.~Cardoso, A.~S.~Miranda, E.~Berti, H.~Witek and V.~T.~Zanchin,
  Phys.\ Rev.\ D {\bf 79}, 064016 (2009)
  doi:10.1103/PhysRevD.79.064016
  [arXiv:0812.1806 [hep-th]].

\bibitem{Konoplya:2017wot}
  R.~A.~Konoplya and Z.~Stuchlík,
  Phys.\ Lett.\ B {\bf 771}, 597 (2017)
  doi:10.1016/j.physletb.2017.06.015
  [arXiv:1705.05928 [gr-qc]].

\bibitem{Dotti:2005sq}
  G.~Dotti and R.~J.~Gleiser,
  Phys.\ Rev.\ D {\bf 72}, 044018 (2005)
  doi:10.1103/PhysRevD.72.044018
  [gr-qc/0503117].

\bibitem{Gleiser:2005ra}
  R.~J.~Gleiser and G.~Dotti,
  Phys.\ Rev.\ D {\bf 72}, 124002 (2005)
  doi:10.1103/PhysRevD.72.124002
  [gr-qc/0510069].

\bibitem{Konoplya:2008ix}
  R.~A.~Konoplya and A.~Zhidenko,
  Phys.\ Rev.\ D {\bf 77}, 104004 (2008)
  doi:10.1103/PhysRevD.77.104004
  [arXiv:0802.0267 [hep-th]].

\bibitem{Takahashi:2011qda}
  Prog.\ Theor.\ Phys.\  {\bf 125}, 1289 (2011)
  doi:10.1143/PTP.125.1289
  [arXiv:1102.1785 [gr-qc]].

\bibitem{Takahashi:2012np}
  T.~Takahashi,
  PTEP {\bf 2013}, 013E02 (2013)
  doi:10.1093/ptep/pts049
  [arXiv:1209.2867 [gr-qc]].

\bibitem{Yoshida:2015vua}
  D.~Yoshida and J.~Soda,
  Phys.\ Rev.\ D {\bf 93}, no. 4, 044024 (2016)
  doi:10.1103/PhysRevD.93.044024
  [arXiv:1512.05865 [gr-qc]].

\bibitem{Cuyubamba:2016cug}
  M.~A.~Cuyubamba, R.~A.~Konoplya and A.~Zhidenko,
  Phys.\ Rev.\ D {\bf 93}, no. 10, 104053 (2016)
  doi:10.1103/PhysRevD.93.104053
  [arXiv:1604.03604 [gr-qc]].

\bibitem{Konoplya:2017ymp}
R.~Konoplya and A.~Zhidenko,
Phys. Rev. D \textbf{95}, no.10, 104005 (2017)
doi:10.1103/PhysRevD.95.104005
[arXiv:1701.01652 [hep-th]].

\bibitem{Konoplya:2017lhs}
  R.~A.~Konoplya and A.~Zhidenko,
  JCAP {\bf 1705}, 050 (2017)
  doi:10.1088/1475-7516/2017/05/050
  [arXiv:1705.01656 [hep-th]].

\bibitem{Konoplya:2017zwo}
  R.~A.~Konoplya and A.~Zhidenko,
  JHEP {\bf 1709}, 139 (2017)
  doi:10.1007/JHEP09(2017)139
  [arXiv:1705.07732 [hep-th]].

\bibitem{Glavan:2019inb}
  D.~Glavan and C.~Lin,
  Phys.\ Rev.\ Lett.\  {\bf 124}, no. 8, 081301 (2020)
  doi:10.1103/PhysRevLett.124.081301
  [arXiv:1905.03601 [gr-qc]].

\bibitem{Konoplya:2020bxa}
  R.~A.~Konoplya and A.~F.~Zinhailo,
  arXiv:2003.01188 [gr-qc].

\bibitem{Guo:2020zmf}
M.~Guo and P.~C.~Li,
Eur. Phys. J. C \textbf{80}, no.6, 588 (2020)
doi:10.1140/epjc/s10052-020-8164-7
[arXiv:2003.02523 [gr-qc]].

\bibitem{Fernandes:2020rpa}
P.~G.~S.~Fernandes,
Phys. Lett. B \textbf{805}, 135468 (2020)
doi:10.1016/j.physletb.2020.135468
[arXiv:2003.05491 [gr-qc]].

\bibitem{Casalino:2020kbt}
  A.~Casalino, A.~Colleaux, M.~Rinaldi and S.~Vicentini,
  arXiv:2003.07068 [gr-qc].

\bibitem{Konoplya:2020qqh}
R.~Konoplya and A.~Zhidenko,
Phys. Rev. D \textbf{101}, no.8, 084038 (2020)
doi:10.1103/PhysRevD.101.084038
[arXiv:2003.07788 [gr-qc]].

\bibitem{Wei:2020ght}
  S.~W.~Wei and Y.~X.~Liu,
  arXiv:2003.07769 [gr-qc].

\bibitem{Kumar:2020owy}
R.~Kumar and S.~G.~Ghosh,
[arXiv:2003.08927 [gr-qc]].

\bibitem{Hegde:2020xlv}
  K.~Hegde, A.~N.~Kumara, C.~L.~A.~Rizwan, A.~K.~M. and M.~S.~Ali,
  arXiv:2003.08778 [gr-qc].

\bibitem{Zhang:2020qew}
  Y.~P.~Zhang, S.~W.~Wei and Y.~X.~Liu,
  arXiv:2003.10960 [gr-qc].

\bibitem{Ghosh:2020syx}
  S.~G.~Ghosh and R.~Kumar,
  arXiv:2003.12291 [gr-qc].

\bibitem{Zhang:2020qam}
  C.~Y.~Zhang, P.~C.~Li and M.~Guo,
  arXiv:2003.13068 [hep-th].

\bibitem{Kumar:2020uyz}
  A.~Kumar and R.~Kumar,
  arXiv:2003.13104 [gr-qc].

\bibitem{HosseiniMansoori:2020yfj}
  S.~A.~Hosseini Mansoori,
  arXiv:2003.13382 [gr-qc].

\bibitem{Wei:2020poh}
S.~W.~Wei and Y.~X.~Liu,
Phys. Rev. D \textbf{101}, no.10, 104018 (2020)
doi:10.1103/PhysRevD.101.104018
[arXiv:2003.14275 [gr-qc]].

\bibitem{Singh:2020nwo}
  D.~V.~Singh, S.~G.~Ghosh and S.~D.~Maharaj,
  arXiv:2003.14136 [gr-qc].

\bibitem{Churilova:2020aca}
  M.~S.~Churilova,
  arXiv:2004.00513 [gr-qc].

\bibitem{Mishra:2020gce}
  A.~K.~Mishra,
  arXiv:2004.01243 [gr-qc].

\bibitem{Heydari-Fard:2020sib}
  M.~Heydari-Fard, M.~Heydari-Fard and H.~R.~Sepangi,
  arXiv:2004.02140 [gr-qc].

\bibitem{Konoplya:2020cbv}
  R.~A.~Konoplya and A.~F.~Zinhailo,
  arXiv:2004.02248 [gr-qc].

\bibitem{Jin:2020emq}
  X.~H.~Jin, Y.~X.~Gao and D.~J.~Liu,
  arXiv:2004.02261 [gr-qc].

\bibitem{Zhang:2020sjh}
  C.~Y.~Zhang, S.~J.~Zhang, P.~C.~Li and M.~Guo,
  arXiv:2004.03141 [gr-qc].

\bibitem{EslamPanah:2020hoj}
B.~Eslam Panah and K.~Jafarzade,
[arXiv:2004.04058 [hep-th]].

\bibitem{NaveenaKumara:2020rmi}
  A.~Naveena Kumara, C.~L.~A.~Rizwan, K.~Hegde, M.~S.~Ali and A.~K.~M,
  arXiv:2004.04521 [gr-qc].

\bibitem{Aragon:2020qdc}
  A.~Aragón, R.~Bécar, P.~A.~González and Y.~Vásquez,
  arXiv:2004.05632 [gr-qc].

\bibitem{Malafarina:2020pvl}
D.~Malafarina, B.~Toshmatov and N.~Dadhich,
Phys. Dark Univ. \textbf{30}, 100598 (2020)
doi:10.1016/j.dark.2020.100598
[arXiv:2004.07089 [gr-qc]].

\bibitem{Yang:2020czk}
  S.~J.~Yang, J.~J.~Wan, J.~Chen, J.~Yang and Y.~Q.~Wang,
  arXiv:2004.07934 [gr-qc].

\bibitem{Cuyubamba:2020moe}
  M.~A.~Cuyubamba,
  arXiv:2004.09025 [gr-qc].

\bibitem{Mahapatra:2020rds}
  S.~Mahapatra,
  arXiv:2004.09214 [gr-qc].

\bibitem{Shu:2020cjw}
  F.~W.~Shu,
  arXiv:2004.09339 [gr-qc].

\bibitem{Casalino:2020pyv}
  A.~Casalino and L.~Sebastiani,
  arXiv:2004.10229 [gr-qc].

\bibitem{Liu:2020evp}
  P.~Liu, C.~Niu and C.~Y.~Zhang,
  arXiv:2004.10620 [gr-qc].

\bibitem{Devi:2020uac}
  S.~Devi, R.~Roy and S.~Chakrabarti,
  arXiv:2004.14935 [gr-qc].

\bibitem{Ma:2020ufk}
  L.~Ma and H.~Lu,
  arXiv:2004.14738 [gr-qc].

\bibitem{Liu:2020yhu}
  P.~Liu, C.~Niu, X.~Wang and C.~Y.~Zhang,
  arXiv:2004.14267 [gr-qc].

\bibitem{Kumar:2020sag}
  R.~Kumar, S.~U.~Islam and S.~G.~Ghosh,
  arXiv:2004.12970 [gr-qc].

\bibitem{Churilova:2020mif}
  M.~S.~Churilova,
  arXiv:2004.14172 [gr-qc].

\bibitem{Ge:2020tid}
  X.~H.~Ge and S.~J.~Sin,
  arXiv:2004.12191 [hep-th].

\bibitem{Zeng:2020dco}
  X.~X.~Zeng, H.~Q.~Zhang and H.~Zhang,
  arXiv:2004.12074 [gr-qc].

\bibitem{Ghosh:2020cob}
S.~G.~Ghosh and S.~D.~Maharaj,
[arXiv:2004.13519 [gr-qc]].

\bibitem{Yang:2020jno}
K.~Yang, B.~M.~Gu, S.~W.~Wei and Y.~X.~Liu,
[arXiv:2004.14468 [gr-qc]].

\bibitem{Gurses:2020ofy}
  M.~Gurses, T.~C.~Sisman and B.~Tekin,
  arXiv:2004.03390 [gr-qc].

\bibitem{Hennigar:2020lsl}
R.~A.~Hennigar, D.~Kubiznak, R.~B.~Mann and C.~Pollack,
[arXiv:2004.09472 [gr-qc]].

\bibitem{Bonifacio:2020vbk}
J.~Bonifacio, K.~Hinterbichler and L.~A.~Johnson,
Phys. Rev. D \textbf{102}, no.2, 024029 (2020)
doi:10.1103/PhysRevD.102.024029
[arXiv:2004.10716 [hep-th]].

\bibitem{Arrechea:2020evj}
J.~Arrechea, A.~Delhom and A.~Jiménez-Cano,
[arXiv:2004.12998 [gr-qc]].

\bibitem{Lu:2020iav}
H.~Lu and Y.~Pang,
[arXiv:2003.11552 [gr-qc]].

\bibitem{Kobayashi:2020wqy}
T.~Kobayashi,
JCAP \textbf{07}, 013 (2020)
doi:10.1088/1475-7516/2020/07/013
[arXiv:2003.12771 [gr-qc]].

\bibitem{Fernandes:2020nbq}
P.~G.~S.~Fernandes, P.~Carrilho, T.~Clifton and D.~J.~Mulryne,
Phys. Rev. D \textbf{102}, no.2, 024025 (2020)
doi:10.1103/PhysRevD.102.024025
[arXiv:2004.08362 [gr-qc]].

\bibitem{Hennigar:2020fkv}
R.~A.~Hennigar, D.~Kubiznak, R.~B.~Mann and C.~Pollack,
[arXiv:2004.12995 [gr-qc]].

\bibitem{Lu:2020mjp}
H.~Lu and P.~Mao,
[arXiv:2004.14400 [hep-th]].

\bibitem{Aoki:2020lig}
K.~Aoki, M.~A.~Gorji and S.~Mukohyama,
[arXiv:2005.03859 [gr-qc]];
[arXiv:2005.08428 [gr-qc]].

\bibitem{Tomozawa:2011gp}
  Y.~Tomozawa,
  arXiv:1107.1424 [gr-qc].
  
\bibitem{Tangherlini:1963bw}
  F.~R.~Tangherlini,
  Nuovo Cim.\  {\bf 27}, 636 (1963).

\bibitem{Myers:1988ze}
  R.~C.~Myers and J.~Z.~Simon,
  Phys.\ Rev.\ D {\bf 38}, 2434 (1988)
  doi:10.1103/PhysRevD.38.2434.

\bibitem{Cognola}
G.~Cognola, R.~Myrzakulov, L.~Sebastiani and S.~Zerbini,
  Phys.\ Rev.\ D {\bf 88}, no. 2, 024006 (2013)
  doi:10.1103/PhysRevD.88.024006
  [arXiv:1304.1878 [gr-qc]].

\bibitem{Cai:2009ua}
  R.~G.~Cai, L.~M.~Cao and N.~Ohta,
  JHEP {\bf 1004}, 082 (2010)
  doi:10.1007/JHEP04(2010)082
  [arXiv:0911.4379 [hep-th]].

\bibitem{Kofinas:2007ns}
  G.~Kofinas and R.~Olea,
  JHEP {\bf 0711}, 069 (2007)
  doi:10.1088/1126-6708/2007/11/069
  [arXiv:0708.0782 [hep-th]].

\bibitem{Takahashi:2010}
  T.~Takahashi and J.~Soda,
  Prog.\ Theor.\ Phys.\  {\bf 124}, 711 (2010)
  doi:10.1143/PTP.124.711
  [arXiv:1008.1618 [gr-qc]];
  Prog.\ Theor.\ Phys.\  {\bf 124}, 911 (2010)
  doi:10.1143/PTP.124.911
  [arXiv:1008.1385 [gr-qc]].

\bibitem{Kodama:2003kk}
H.~Kodama and A.~Ishibashi,
Prog. Theor. Phys. \textbf{111}, 29-73 (2004)
doi:10.1143/PTP.111.29
[arXiv:hep-th/0308128 [hep-th]].

\end{thebibliography}
\end{document}